\documentclass[aps,prd,showpacs,amssymb,twocolumn,nofootinbib]
{revtex4}

\begin{document}

\title{Entropy and temperature of black holes in a gravity's rainbow}

\author{Pablo \surname{Gal\'an}}
\affiliation{Instituto de Estructura de la Materia, CSIC, Serrano
121, 28006 Madrid, Spain}

\author{Guillermo A. \surname{Mena Marug\'an}}
\affiliation{Instituto de Estructura de la Materia, CSIC, Serrano
121, 28006 Madrid, Spain}

\begin{abstract}
The linear relation between the entropy and area of a black hole can
be derived from the Heisenberg principle, the energy-momentum
dispersion relation of special relativity, and general
considerations about black holes. There exist results in quantum
gravity and related contexts suggesting the modification of the
usual dispersion relation and uncertainty principle. One of these
contexts is the gravity's rainbow formalism. We analyze the
consequences of such a modification for black hole thermodynamics
from the perspective of two distinct rainbow realizations built from
doubly special relativity. One is the proposal of Magueijo and
Smolin and the other is based on a canonical implementation of
doubly special relativity put forward recently by the authors. In
these scenarios, we obtain modified expressions for the entropy and
temperature of black holes. We show that, for a family of doubly
special relativity theories satisfying certain properties, the
temperature can vanish in the limit of zero black hole mass. For the
Magueijo and Smolin proposal, this is only possible for some
restricted class of models with bounded energy and unbounded
momentum. With the proposal of a canonical implementation, on the
other hand, the temperature may vanish for more general theories; in
particular, the momentum may also be bounded, with bounded or
unbounded energy. This opens new possibilities for the outcome of
black hole evaporation in the framework of a gravity's rainbow.
\end{abstract}


\maketitle
\renewcommand{\theequation}{\arabic{section}.\arabic{equation}}

\section{Introduction}

More than thirty years ago, Bekenstein argued that the entropy of a
black hole is a linear function of the area of its event horizon
\cite{bek}. He also proposed a value for the proportionality
constant, deduced from a semiclassical calculation of the minimum
increase in the area of a black hole when it absorbs a particle.
Bekenstein's line of reasoning can in fact be generalized by
considering the quantum nature of the particle and taking then into
account the uncertainty principle and the energy-momentum dispersion
relation \cite{hod}. This generalized argument leads essentially to
the same conclusion about the linearity of the entropy with respect
to the black hole area.

Subsequent works in different formalisms for quantum gravity
(specially in string theory and loop quantum gravity) have not only
provided an explanation to Bekenstein's result
\cite{bekstring,bekloop}, but also revealed that the linear behavior
of the entropy should be modified by a leading order correction that
is logarithmic for large areas \cite{logstring,logloop}. Similar
results have been derived also by considering general properties of
black holes \cite{loggeneral}.

As we have commented, the uncertainty principle and the dispersion
relation play a key role in the quantum generalization of the
Bekenstein argument. If one accepts the possibility that either (or
both) of these elements suffers modifications, one will deduce a
different result, with the appearance of terms additional to that
proportional to the area. The effect of changes in the uncertainty
principle was considered in Refs. \cite{loggeneral,chen,medved},
whereas a more general analysis including modified dispersion
relations was recently presented by Amelino-Camelia {\it et al}
\cite{amelinoBH1,amelinoBH2}, and by Ling and collaborators
\cite{ling1,ling2}. In particular, with a suitable modification of
the dispersion relation and/or the uncertainty principle, a
logarithmic term can be obtained.

Modifications to the standard dispersion relations and uncertainty
principle arise indeed in several approaches to quantum gravity. For
instance, one can find modified dispersion relations in quantum
descriptions of spacetime that involve a discrete geometry, such as
loop quantum gravity \cite{MDRloop,MDRloop2,MDRloop3}, and in
schemes that adopt a noncommutative spacetime geometry
\cite{MDRnoconm}. On the other hand, generalized uncertainty
principles have been derived in the context of string theory
\cite{GUPstring,GUPstring2,GUPstring3}, in descriptions using
noncommutative geometry \cite{GUPnoconm}, and in other kinds of
analysis based on general considerations about the interplay between
quantum mechanics and gravity \cite{GUP0}.

Modified dispersion relations have also been studied from a
phenomenological point of view, owing to the increasing interest in
discussing their observational consequences
\cite{fenom,fenom2,fenom3,fenom4,fenom5}. In addition, one
encounters deformed dispersion relations in the so-called doubly
special relativity (DSR) theories
\cite{DSRi,DSRii,DSR1,DSR1b,DSR2,DSR2b,DSR12}. The initial
motivation for these theories was to solve the apparent
inconsistency that exists between the relativity principle and the
emergence of a fundamental scale (Planck scale), suggested by most
approaches to quantum gravity \cite{Luis}. The compatibility is
regained in DSR theories by allowing a nonlinear action of the
Lorentz symmetry. With this modification, not only energy and
momentum cease to obey the standard dispersion relation, but also
the conventional uncertainty relations (involving energy-momentum
and spacetime) are generically affected. In this way, the framework
of DSR theories permits to deal simultaneously with both types of
modifications in a quite general manner. Because of this reason, we
will concentrate our attention on this framework from now on.
Specifically, we will consider two different proposals to implement
the consequences of DSR on the spacetime geometry. One of these
proposals is the gravity's rainbow put forward by Magueijo and
Smolin (MS) \cite{rainbow}. The other is based on a canonical
implementation of the DSR theories recently suggested by the authors
\cite{nuestro,nuestro2}. Actually, this second proposal can be
regarded also as a sort of gravity's rainbow formalism inasmuch as
it leads to spacetime metrics with an explicit dependence on the
energy-momentum (corresponding to the test particles employed by the
observer \cite{rainbow}).

The aim of this work is to discuss the effect that the modification
of the dispersion relations and the uncertainty principle entail on
black hole thermodynamics in the context of a gravity's rainbow.
Many of the ideas of this discussion are inspired by those proposed
in Ref. \cite{amelinoBH2} (see also
\cite{loggeneral,chen,medved,amelinoBH1}), which provides the first
detailed study of the combined effects of these types of
modifications, and in the further elaboration of the arguments of
that paper presented in Refs. \cite{ling1,ling2}. In particular,
Ref. \cite{ling2} is the first discussion of the changes expected
for black hole thermodynamics in a gravity's rainbow. Merit for the
initial ideas must be granted to those works, though the
contributions of our analysis are manifold: we depurate and
systematize the arguments of those references for their application
to gravity's rainbow schemes, we extend the conclusions to more
general families of DSR theories (allowing not just a deformation of
the energy, but also a generic deformation of the momentum), and we
generalize the analysis to a gravity's rainbow formalism that
differs from the MS one, proving that this alternative candidate
leads to a thermodynamics with specially appealing physical
properties.

The paper is organized as follows. Section II reviews the (extended)
Bekenstein argument including also quantum considerations. In Sec.
III we summarize some aspects of DSR theories. We comment the
gravity's rainbow formalism introduced by Magueijo and Smolin and
the one corresponding to our proposal for a canonical implementation
of DSR. For each of these formalisms, we derive new spacetime
coordinates, referred to as physical. In Sec. IV we make some
considerations about the quantum description of the system and its
``relation'' to the two studied gravity's rainbow formalisms. We
also obtain a modified lower bound for the change of black hole
area. We deduce modified expressions for the black hole entropy in
Sec. V and for the temperature in Sec. VI. Our conclusions are
contained in Sec. VII. Finally, an appendix is added. We will use a
set of units in which $\hbar=c=1$, so that $L_P=E_P^{-1}=\sqrt{G}$
(with $\hbar$ being the Planck constant, $c$ the speed of light, $G$
the gravitational constant, $L_P$ the Planck length, and $E_P$ the
Planck energy).

\section{Extended Bekenstein Argument}

In order to calculate the proportionality constant in the linear
entropy-area relation, Bekenstein used a semiclassical argument
\cite{bek} that generalizes the process of a particle falling into a
black hole discussed previously by Christodoulou \cite{christ}. As a
part of his argument, Bekenstein calculated the minimum growth of
area that a black hole undergoes when it swallows a (neutral)
classical particle with energy $E$ and proper radius $L$
\cite{note1} that crosses the horizon falling freely from a turning
point in its orbit \cite{bek,hod}. He concluded that $\Delta
A\geq8\pi L_P^2 EL.$ Therefore, the area increase displays a
fundamental lower bound that does not depend on the black hole
properties.

In order to extend this analysis to the case of a quantum particle,
one has to regard the radius of the particle as the uncertainty in
its position, $\Delta x$, and introduce an appropriate correction
factor in the coefficient of the above expression for $\Delta A.$ We
will generically designate this corrected coefficient by $a$. One
then has
\begin{equation}\label{area}\Delta A\geq a E\Delta x.\end{equation}
Using the dispersion relation of special relativity and the usual
uncertainty principle, one gets $E\geq 1/\Delta x$
\cite{amelinoBH1,amelinoBH2} and, therefore, $\Delta A\geq a.$

On the other hand, Bekenstein also pointed out the existence of a
universal upper bound on the entropy-to-energy ratio \cite{cotaS/E},
$S/E\leq 2\pi L$, where $L$ is the effective radius of the system.
So, if a quantum system with entropy $S_{mat}$ enters a black hole,
the change of the ordinary matter entropy in the black hole exterior
satisfies
\begin{equation}-\Delta S_{mat}\leq b E\Delta x,\end{equation}
where we have denoted the corresponding proportionality constant by
$b$. This bound, together with Eq. (\ref{area}), implies that $(b/a)
\,\Delta A + \Delta S_{mat}\geq0$. This inequality can be viewed as
a generalized second law of thermodynamics, establishing that the
first term in the expression represents the change in the black hole
entropy, $\Delta S_{BH}$ \cite{bek}. By adjusting properly the
coefficient $b/a$, one arrives at the well known result:
\begin{equation}\label{S0}S_{BH}=\frac{A}{4L_P^2}.\end{equation}

For a Schwarzschild black hole, case to which we restrict our
attention from now on for simplicity, the associated temperature can
be deduced then by employing the definition $T_{BH}^{-1}=\partial_m
S_{BH}$ \cite{bek}, where $m$ is the mass of the black hole. If one
considers the usual relation $A=16\pi L_P^4 m^2,$ one gets
\begin{equation}\label{T0}T_{BH}=\frac{E_P^2}{8\pi m},\end{equation}
which reproduces the temperature that Hawking obtained in a
different way \cite{hawking}.

The uncertainty principle and the dispersion relation have a key
role in the derivation that we have presented. If either or both of
them experienced modifications, as it happens in the various
frameworks that we have commented
\cite{MDRloop,MDRloop2,MDRloop3,MDRnoconm,GUPstring,
GUPstring2,GUPstring3,GUPnoconm,GUP0}, the linearity in the
expression of the entropy would also be modified. In the rest of the
paper, we will analyze the implications that the modifications
introduced by DSR theories have on black hole thermodynamics.

\section{Doubly Special Relativity and Gravity's Rainbow}
\setcounter{equation}{0}

DSR theories are characterized by the inclusion of a Lorentz
invariant energy and/or momentum scale, in addition to the
fundamental scale provided by the speed of light
\cite{DSRi,DSRii,DSR1,DSR1b,DSR2,DSR2b,DSR12}. The invariance of
this new scale (supposed to be related to the Planck scale) is
possible thanks to a nonlinear action of the Lorentz group in
momentum space. A realization of this kind is obtained via an
invertible nonlinear map $U$ between the physical energy-momentum
$P_{a}:=(-E,p_i)$ and the original energy and momentum variables of
standard relativity (in Minkowski space)
$\Pi_{a}:=(-\epsilon,\Pi_i)$, which are viewed as auxiliary
variables \cite{judes} (lowercase Latin indices from the beginning
and the middle of the alphabet represent Lorentz and flat spatial
indices, respectively). Imposing that the action of rotations is not
modified, the nonlinear map $U$ is totally determined by two scalar
functions $g$ and $f$ \cite{DSR2b,DSRposition}. Following a notation
similar to that of Refs. \cite{nuestro,nuestro2}, the map $U$ can be
expressed
\begin{equation}\label{momenta}
P_a = U^{-1}(\Pi_a) \Rightarrow \left\{ \begin{array}{l} E =
g(\epsilon,\Pi),\\
$$\displaystyle p_i =f(\epsilon,\Pi)\frac{\Pi_i}{\Pi}.$$
\end{array} \right.\end{equation}
Here, $\Pi$ denotes the magnitude of the auxiliary momentum. We get
different DSR theories depending on the choice of functions $f$ and
$g$. On the other hand, to recover the standard linear action of the
Lorentz group in the limit of small energies and momenta compared to
the scale of the DSR theory, one must impose that the functions
$(g,f)$ tend to the identity [i.e., behave like $(\epsilon,\Pi)$] in
that limit.

In order to determine the corresponding transformation rules in
position space and the deformed spacetime geometry, there exist
different proposals in the literature
\cite{DSRposition,DSRposition2,DSRposition3,rainbow,nuestro,nuestro2}.
In this paper, we will focus on the MS proposal of a gravity's
rainbow \cite{rainbow} and on a variant of it based on a canonical
implementation of DSR \cite{nuestro,nuestro2}.

The MS proposal rests on the requirement that the contraction
between the energy-momentum and an infinitesimal spacetime
displacement be a linear invariant in DSR. In contrast, our proposal
demands the invariance of the symplectic form ${\bf d} q^a \wedge
{\bf d} \Pi_a$, where $q^a$ represents the (asymptotically) flat
spacetime coordinates, that we will refer to as auxiliary. Both
proposals lead to geometries that depend explicitly on the energy
and momentum of the system (namely, the test particle used by the
observer \cite{rainbow}). This fact explains the name {\it gravity's
rainbow} \cite{rainbow} given to this type of formalisms. Unlike
what happens with the MS proposal, ours leads to modified spacetime
coordinates $x^a$ that are conjugate to the physical energy-momentum
$P_a$. Namely, the relation between $(q^a,\Pi_a)$ and $(x^a,P_a)$ is
a canonical transformation. Similar canonical proposals for the
implementation of DSR have been suggested by other authors
\cite{DSRposition3,hinterleitner}.

We will concentrate our discussion on a family of DSR theories that,
without being completely generic, is in fact rather general (at
least compared with the cases studied so far in connection with
black hole thermodynamics). In these theories, the physical energy
depends only on the auxiliary one, i.e. $E=g(\epsilon)$. In
addition, we require the ratio of the physical and auxiliary momenta
to be well defined when the latter of these momenta tends to zero.
This is a minor restriction on $f$ \cite{note1b}, since in any case
it must be approximately equal to $\Pi$ for small energies and
momenta. Our condition guarantees that $f(\epsilon,\Pi)$ vanishes
when so does the auxiliary momentum. Finally, since we are only
considering spherically symmetric black holes for simplicity, we
will impose spherical symmetry also on (the test particle) phase
space, restricting to physical momenta that are parallel or
antiparallel to $x^i$. As a consequence, $\varepsilon^{i} _{jk}p_i
dx^j=0$ (for given momentum), where $\varepsilon_{ijk}$ denotes the
Levi-Civita symbol. Since $p_i/p=\Pi_i/\Pi$, this condition can be
rewritten as $dx^i=dx^j \Pi_j\Pi^i/\Pi^2$ (with the usual sum
convention in repeated indices).

In these circumstances, one gets the following scaling with our
proposal of a canonical implementation of DSR (see \cite{nuestro}):
\begin{eqnarray}\label{GM}
dq^0&=&\partial_{\epsilon}g\left(dx^0\pm\frac{\partial_{\epsilon}f}
{\partial_{\epsilon}g}dx\right):=\partial_{\epsilon}g
d\tilde{x}^0_{\pm}, \nonumber\\
dq^i&=&\partial_{\Pi}f dx^i.
\end{eqnarray} Here $x=\sqrt{x^i x^i}$, and $\tilde{x}^0_{\pm}$ is a
new coordinate that, although not canonically conjugate to the
energy (in the sense that $\{\tilde{x}^0_{\pm},x^i\}$ is not a set
conjugate to the physical energy-momentum), differs from the
canonical time only in a shift that is constant in spacetime.

Following the MS proposal, on the other hand, one arrives at
\cite{rainbow}:
\begin{equation}\label{MS}
dq^0=\frac{g}{\epsilon}dx^0,\quad \quad dq^i=\frac{f}{\Pi}dx^i.
\end{equation}
Therefore we see that, in the two considered cases, the effect on
the geometry consists essentially of two independent scalings: a
conformal transformation of the spatial components and a time
dilation, both of them constant in spacetime. For instance, one can
obtain the modified Schwarzschild solution for the gravity's rainbow
following the steps explained in detail in Ref. \cite{rainbow}. This
solution reproduces formally the familiar one for general relativity
(with a suitable identification of coordinates) except for the
commented scaling of the spatial metric and the diagonal time
component.

We can analyze simultaneously the two gravity's rainbow formalisms
by denoting the corresponding scale factors with the abstract
notation $G(\epsilon)$ and $F(\epsilon,\Pi)$:
\begin{equation}\label{GRS}
dq^0=G(\epsilon)d\tilde{x}^0_{\pm},\quad \quad
dq^i=F(\epsilon,\Pi)dx^i,
\end{equation}
with $\tilde{x}^0_{\pm}$ designating $x^0$ for the MS proposal. Note
that the time and spatial scale factors are given in one case by the
partial derivatives of the functions $g$ and $f$ with respect to the
auxiliary energy and momentum, respectively, whereas in the other
case they are simply the ratios of those quantities, namely:
\begin{eqnarray}
G(\epsilon)&:=&\left\{\begin{array}{ll}
$$\displaystyle\frac{g(\epsilon)}{\epsilon}$$ & {\rm
MS\;proposal,} \\ \partial_{\epsilon}g(\epsilon)\quad\quad & {\rm
Canonical\;\;proposal,}\end{array}\right.\label{funG}\\
F(\epsilon,\Pi)&:=&\left\{\begin{array}{ll}
$$\displaystyle\frac{f(\epsilon,\Pi)}{\Pi}$$ &
{\rm MS\;proposal,}\\
\partial_{\Pi}f(\epsilon,\Pi) \;\;& {\rm Canonical
\;\;proposal.}\end{array}\right.\label{funF}\end{eqnarray} By
canonical proposal, we understand here our proposal for a canonical
implementation of DSR.

Expressions (\ref{momenta}) and (\ref{GRS}) lead to deformations of
the dispersion relation and to generalized uncertainty principles
(because the commutators of the momentum with the auxiliary spatial
coordinates vary). In this way, the gravity's rainbow formalisms
incorporate in fact the two types of modifications whose
consequences for black holes we want to discuss.

\section{Modified Bound on the Change of Black Hole Area}
\setcounter{equation}{0}
\subsection{Quantum description of the system}

Expression (\ref{area}) provides a lower bound for the increase of
black hole area in general relativity. The magnitude $E$ is the
energy of the particle which is going to be absorbed, measured at
infinity in the asymptotically flat spacetime, and $\Delta x$ is the
uncertainty in the position of the particle. When relativity is
modified, it seems natural to assume that the bound continues to
apply with $E$ being the energy measured by an asymptotic observer
\footnote{See nonetheless Subsec. IV B concerning the case of a
perturbative quantization.} and $\Delta x$ the position uncertainty.
However, in DSR, $E$ and $x$ no longer correspond to the standard
energy and position variables for (asymptotically) flat spacetime:
they are the variables that we have called physical and transform
(in the asymptotic region) according to a nonlinear action of the
Lorentz group. In this way, the expression for the area increase
incorporates modifications with respect to Eq. (\ref{area}) arising
from the DSR deformation.

In the following, we will call $\Delta A$ the change of area
obtained for DSR, whereas $\Delta A_0:=a \epsilon \Delta q$ denotes
the undistorted lower bound for standard general relativity [see Eq.
(\ref{area})]. In order to relate these two quantities, we have to
take into account first the kind of quantum description adopted. We
can consider two possibilities. In one case, the quantization of the
system is carried out choosing as time parameter and position
variables the auxiliary coordinates corresponding to (the
asymptotically) flat spacetime, which can be seen as a background.
This is the typical philosophy of a perturbative approach. In the
other case, on the contrary, the quantization is constructed in
terms of the physical time and position variables. For this reason,
we will refer to these two types of descriptions as perturbative and
nonperturbative quantization, respectively \cite{nuestro,nuestro2}.

More specifically, for the perturbative quantization a complete set
of elementary variables are the auxiliary variables $(q^i, \Pi_i)$,
while the auxiliary time $q^0$ plays the role of evolution
parameter. Hence, the uncertainty principle corresponding to this
perturbative description is $\Delta q \Delta\Pi \geq1/2$ (with
$q=\sqrt{q^i q^i}$) \cite{note2}. On the other hand, the
nonperturbative quantization can be built by regarding the physical
time $x^0$ as the evolution parameter and the canonically conjugate
physical variables $(x^i,p_i)$ as the elementary ones. The
uncertainty principle for this nonperturbative description is then
$\Delta x\Delta p\geq1/2$. Our next task consists in expressing the
quantities that appear in $\Delta A$ in terms of the elementary
variables that correspond to each of these types of quantization and
apply the uncertainty principle associated with them.

\subsection{Bound on the change of area}

For the case of the nonperturbative quantization, the bound on the
change of area for DSR can be expressed
\begin{equation}
\Delta A \geq a E\Delta x =  \frac{E}{\epsilon} \frac{\Delta
\Pi}{\Delta p} \frac{\Delta x \Delta p}{\Delta q \Delta \Pi} \Delta
A_0.\end{equation} In accord with the absorption process described
in Sec. II, we restrict to particles that are away from rest an
amount $\Delta \Pi$, interpretable as the uncertainty that affects
their momenta. Employing Eq. (\ref{momenta}) and the fact that the
function $f$ vanishes when so does $\Pi$, we conclude
\begin{equation}\label{areaMS}  \Delta A\geq
\frac{g/\epsilon}{f/\Delta\Pi} \widehat{\Delta A}_0,
\end{equation}
where we have defined
\begin{equation}
\widehat{\Delta A}_0:=  \frac{\Delta x \Delta p}{\Delta q \Delta
\Pi} \Delta A_0.
\end{equation}
The factor $\Delta x \Delta p/(\Delta q \Delta \Pi)$ in
$\widehat{\Delta A}_0$ compensates the change of basic uncertainty
relations with respect to those for flat spacetime in the Bekenstein
argument. Taking into account this change, the line of reasoning of
Sec. II would lead to the result $(b/a)\widehat{\Delta A}_0+\Delta
S_{mat}\geq 0$. Note that the ratio of the scale factors encountered
in Eq. (\ref{MS}) appears now in the right hand side of expression
(\ref{areaMS}). According to this fact, the MS gravity's rainbow
formalism seems to match with the nonperturbative description.

Let us consider now the perturbative description. Recalling Eq.
(\ref{GM}) and treating $\Delta x$ and $\Delta q$ as spatial
distances, one gets $\Delta x=\Delta q/\partial_\Pi f$. In addition,
if the auxiliary energy $\epsilon$ is viewed quantum mechanically as
the generator of time translations in the evolution parameter $q^0$
and one identifies the quantity $E$ in $\Delta A$ with the
corresponding generator of translations in $x^0$ obtained with a
straight application of the chain rule, rather than with the exact
physical energy, one concludes that the role of $E$ must be played
by $\epsilon\partial_{\epsilon}g$. This same assignation was in fact
made in Ref. \cite{amelinoBH2} when studying the leading order
corrections caused by modified dispersion relations in black hole
thermodynamics. With these considerations,
\begin{equation}\label{areaGM}
\Delta A \geq \partial_{\epsilon}g a \epsilon  \Delta x =
\frac{\partial_\epsilon g}{\partial_\Pi f}\Delta A_0.\end{equation}

In the above expression, the multiplicative factor is the ratio of
the scale factors obtained in Eq. (\ref{GM}). Thus, the same type of
connection that seems to exist between the nonperturbative quantum
description and the MS proposal appears now between the perturbative
description and our proposal for the canonical implementation of
DSR. In both cases, the bound on the change of area for general
relativity is corrected by a factor that depends on the auxiliary
energy and momentum of the particle. In the case of the
nonperturbative quantization, there is an additional modification
coming from the change in the uncertainty relations, which has been
absorbed in the definition of $\widehat{\Delta A}_0$. In the regime
of low energies, the functions $g$ and $f$ tend to the identity and
the physical and auxiliary variables coincide, so that the standard
result is recovered.

We can deal simultaneously with formulas (\ref{areaMS}) and
(\ref{areaGM}) by employing the notation introduced in Eqs.
(\ref{funG}) and (\ref{funF}) and calling both $\widehat{\Delta
A}_0$ in the nonperturbative description and $\Delta A_0$ in the
perturbative one by $\overline{\Delta A}_0$. We can then write
\begin{equation}\label{area2} \Delta A
\geq \frac{G(\epsilon)}{F(\epsilon,\Delta\Pi)}\overline{\Delta
A}_0:=H(\epsilon,\Delta\Pi)\overline{\Delta A}_0.\end{equation}
Explicitly
\begin{equation}H(\epsilon,\Pi)=\left\{\begin{array}{ll}
$$\displaystyle\frac{\Pi g(\epsilon)}{\epsilon f(\epsilon,\Pi)}$$
& {\rm MS\;proposal,} \\
$$\displaystyle\frac{\partial_{\epsilon}g(\epsilon)}
{\partial_{\Pi}f(\epsilon,\Pi)}$$ \;& {\rm
Canonical\;\;proposal.}\end{array}\right.\nonumber\end{equation} We
will use this relation to derive modified expressions for the black
hole entropy and temperature in the two considered gravity's rainbow
formalisms. In principle, this inequality is valid for all possible
values of the energy-momentum of the particle absorbed by the black
hole. At least for the simple case of a modified Schwarzschild black
hole, it turns out that when the function $H$ satisfies certain
conditions, the set of inequalities obtained with different energies
and momenta amounts just to a single inequality. Furthermore, the
factor arising from $H$ in that inequality depends only on the area
radius $r_s=\sqrt{A/(4\pi)}$ (i.e., the Schwarzschild radius in
general relativity) \cite{note3}. The conditions on $H$ follow from
the next arguments.

First, since the auxiliary energy and momentum satisfy the usual
dispersion relation of special relativity, $\epsilon\geq \Delta \Pi$
(with the equality attainable for massless particles). Therefore, if
$H(\epsilon,\Delta \Pi)$ is an increasing function of the variable
$\Delta\Pi$, we can maximize it to $H(\epsilon,\epsilon)$. In
addition, $2\pi/\epsilon$ cannot exceed the diameter of the black
hole, $2r_s$, because otherwise the particle would be scattered
instead of absorbed (see also Ref. \cite{amelinoBH2}). We notice
that $2\pi/\epsilon$ is the wavelength in the case of a massless
particle, and it is smaller or equal than the Compton wavelength if
the particle is massive. So, $\epsilon\geq \pi/r_s$. If
$H(\epsilon,\epsilon)$ is a decreasing function of $\epsilon$, it
then reaches its maximum at $\pi/r_s$.

In this situation, it is easy to see that expression (\ref{area2})
is satisfied for all allowed values of the auxiliary energy-momentum
if and only if it is satisfied for $\epsilon=\Delta\Pi=\pi/r_s$
\cite{note4}, namely
\begin{equation}\label{areacota}\Delta A\geq\ H\left(\frac{\pi}{r_s},
\frac{\pi}{r_s}\right) \overline{\Delta A}_0.\end{equation} In the
next section, we will use this inequality as the key element to
derive modified expressions for the entropy and temperature of the
black hole. Let us check now if the conditions demanded on the
function $H$ are fulfilled in some of the DSR models that are more
often found in the literature (see the Appendix for details).

The first of these DSR models can be considered the prototype of the
DSR2 family \cite{DSR2,DSR2b} (i.e., theories with bounded physical
energy and momentum). It is seen in the Appendix that
$H(\epsilon,\Delta\Pi)=g(\epsilon)\Delta\Pi/ [\epsilon
f(\epsilon,\Delta\Pi)]=1$ in this model for the MS proposal.
Therefore, both requirements about the function $H$ are trivially
satisfied, but there is no modification of the thermodynamics. For
our canonical implementation, on the other hand,
$H(\epsilon,\Delta\Pi)=\partial_\epsilon g/\partial_\Pi f=1/(1+
\lambda\epsilon)$. Since this function is constant in $\Delta\Pi$
and decreasing in $\epsilon$, the conditions are satisfied.

Our next example is a DSR analogue of the Einstein-Rosen
gravitational waves \cite{ER}, which is of the DSR3 class (i.e.,
only the physical energy is bounded). It is shown in the Appendix
that in this case one gets for the MS proposal
$H(\epsilon,\Delta\Pi)=(1-e^{-\lambda\epsilon})/ (\lambda\epsilon)$,
and for our proposal $H(\epsilon,\Delta\Pi)=e^{-\lambda\epsilon}.$
Both functions are independent of $\Delta\Pi$ and decreasing in
$\epsilon$. Hence the requirements are fulfilled.

Finally, we consider the model of DSR1 class (i.e., with bounded
physical momentum) introduced in Refs. \cite{DSRi,DSRii}. In fact,
our analysis cannot be applied in this model because the physical
energy depends on the auxiliary momentum, $E=g(\epsilon,\Delta\Pi)$,
unless one restricts all considerations to fixed Casimir invariant
$\epsilon^2-\Delta\Pi^2$. For the MS proposal, given the complexity
of the involved functions, we study exclusively the case of massless
particles ($\epsilon=\Delta\Pi$), for which we show in the Appendix
that $H=(1+\lambda\epsilon)\ln(1+\lambda\epsilon)/
(\lambda\epsilon)$. This is an increasing function of $\epsilon$.
Thus, the conditions are not satisfied. For our proposal, on the
other hand, one obtains $H=1$ with any fixed value of the Casimir
invariant, and hence the thermodynamics remains unaltered.

\section{Black Hole Entropy} \setcounter{equation}{0}

Let us discuss now the modification of the entropy-area relation
(\ref{S0}). As we have seen, in standard general relativity we have
$(b/a) \overline{\Delta A}_0+\Delta S_{mat}\geq 0$ (once the proper
basic uncertainty relations have been taken into account). In the
passage to deformed relativity, the change of black hole area
satisfies relation (\ref{areacota}) provided that the function $H$
fulfills certain conditions. In this passage, the entropy of
ordinary matter is not modified (assuming that one has already
adopted a correct quantum description), because it simply reflects
the number of physical degrees of freedom of the system. Combining
this information, one concludes
\begin{equation}
\frac{b}{a} \frac{\Delta A}{H\left(\frac{\pi}{r_s},
\frac{\pi}{r_s}\right)} +\Delta S_{mat}\geq 0.
\end{equation}

If one accepts the reasonable hypothesis that the (Schwarzschild)
black hole entropy is a function of its area only, the above
relation can be understood as a modified generalized second law,
with a natural identification of the change in black hole entropy:
\begin{equation}\label{entropy}\Delta S_{BH}=\frac{b}{a}
\frac{\Delta A} {H\left(\frac{\pi}{r_s},
\frac{\pi}{r_s}\right)}.\end{equation} Notice that, to recover the
Bekenstein-Hawking law (\ref{S0}) for large black holes
($r_s\rightarrow\infty$), limit in which $H$ tends to the unity, the
constant $b/a$ must be fixed equal to the usual factor $1/(4L_P^2)$.

Finally, in order to obtain the functional form of the entropy with
the area, we substitute $A=4\pi r_s^2$. Integrating Eq.
(\ref{entropy}) we get \cite{ling1,ling2}
\begin{equation}\label{entropy2} S_{BH}(A)=\frac{1}{4L_P^2}
\int_{A_1}^{A} \frac{d \tilde{A}}{H\left(\sqrt{4\pi^3/\tilde{A}},
\sqrt{4\pi^3/\tilde{A}}\right)}.\end{equation} Here, $A_1$ is a
reference area where the entropy is fixed to vanish. It is natural
to choose it equal to zero, but we note that the integral might then
diverge. In that case, a nonzero area (e.g. the Planck area) should
be given as the reference. The convergence of the integral for
$A_1=0$ is ensured at least for those theories in which one can find
a constant $\delta>0$ so that $\lim_{r_s\rightarrow 0}
r_s^{2-\delta}/H(\pi/r_s,\pi/r_s)=0$.

Let us study now the behavior of the entropy for large values of the
black hole area. With this aim, we expand the function $1/H$ around
zero (in both of its arguments) and keep only up to quadratic terms.
Remembering that $H(0,0)=1$, we obtain after integration (up to an
irrelevant additive constant)
\begin{equation}
S_{BH}\approx\frac{1}{4L_P^2}\left[A+ 4 C_1\sqrt{\pi^3 A}+2\pi^3
C_2\ln{\frac{A}{L_P^2}} \right],\end{equation}with
\begin{eqnarray}
C_1 &=&-(\partial_\epsilon H+\partial_\Pi H)\big|_0,\\
C_2 &=&\left[-\partial_{\epsilon}^2 H
-\partial_{\Pi}^2H-2\partial_{\Pi}\partial_{\epsilon}H+2
(\partial_\epsilon H+\partial_\Pi H)^2\right]\big|_0.\nonumber
\end{eqnarray} Here, the symbol $|_0$ denotes evaluation at vanishing
arguments.

If one imposes that the leading order correction to the
Bekenstein-Hawking law be logarithmic, in agreement with several
analyses in the literature \cite{logstring,logloop,loggeneral}, the
constant coefficient $C_1$ must vanish. This can be understood as a
restriction on the allowed DSR theories. In that case, the
coefficient $C_2$ becomes
\begin{equation}
C_2 =- \left[\partial_{\epsilon}^2 H +\partial_{\Pi}^2
H+2\partial_{\Pi}\partial_{\epsilon}H\right]\big|_0.
\end{equation}

\section{Black Hole Temperature}
\setcounter{equation}{0}
\subsection{Derivation from the entropy}

We turn now to analyze the modification of the black hole
temperature. With this purpose, we associate a mass $m=E_P^2r_s/2$
to the black hole. This expression reproduces the definition of
Schwarzschild mass in general relativity and is recovered in the
gravity's rainbow formalisms \cite{rainbow}. It is worth pointing
out that the mass $m$ runs in principle from zero to infinity (if so
does $A$). From the definition $T_{BH}^{-1}:=\partial_{m} S_{BH}$
and Eq. (\ref{entropy}) [with $b/a=1/(4L_P^2)$], one arrives then at
the modified temperature
\begin{equation}\label{temperature}T_{BH}=H\left(4\pi^2 T_0,
4\pi^2 T_0\right) T_0.
\end{equation}
Here, $T_0=E_P^2/(8\pi m)$ represents the Hawking temperature in
standard general relativity.

The Hawking temperature tends to zero as $m\rightarrow\infty$ (or
equivalently as $r_s\rightarrow\infty$). So, in general relativity
the black hole radiates with a negligible temperature when its mass
is very large. On the contrary, $T_0$ diverges when $m$ (and $r_s$)
approaches zero. As a consequence, in general relativity, the amount
of radiation emitted by a tiny black hole is enormous. The
evaporation accelerates explosively when the mass becomes small, in
the final stages of the black hole lifetime. We want to explore
whether the modifications that arise in the context of a gravity's
rainbow can significantly change this behavior of the temperature.
Specifically, we want to investigate whether the modified
temperature can vanish in the limit of zero black hole mass. This
would open the possibility that the black hole evaporation
eventually stops or takes an infinite time, providing a radically
different scenario for the resolution of the information paradox.

Let us call $z:=4\pi^2 T_0= \pi E_P^2/(2m)$. Employing Eq.
(\ref{areaMS}), one obtains for the MS proposal
\begin{equation}T_{BH}=\frac{1}{4\pi^2}\frac{g(z)}{f(z,z)}z.
\end{equation}
We remember that $g$ is a positive and increasing function, because
it is approximately the identity at low energies and is invertible.
For the temperature $T_{BH}$ to vanish in the zero mass limit, it is
then necessary (though not sufficient) that
$\lim_{z\rightarrow\infty}f(z,z)/z=\infty.$ One can realize that
this precludes the existence of an invariant momentum scale.
Therefore, the temperature cannot vanish for theories that belong to
the DSR1 and DSR2 classes.

On the other hand, from Eq. (\ref{areaGM}) the modified temperature
for our proposal is
\begin{equation}T_{BH}=\frac{1}{4\pi^2}\frac{\partial_\epsilon g(z)}
{\partial_\Pi f(z,z)}z.\end{equation} In theories of the DSR2 and
DSR3 classes, $\partial_\epsilon g(z)$ tends to zero at infinity. If
the decrease of this derivative dominates over the possible increase
of $z/\partial_\Pi f(z,z)$, the temperature vanishes for zero mass.
In the DSR1 class, $\partial_\epsilon g(z)$ does not tend to zero.
So, one must necessarily have $\lim_{z\rightarrow\infty}\partial_\Pi
f(z,z)/z=\infty$. Since the existence of a bound on the physical
momentum implies only that $\lim_{z\rightarrow\infty}f(z,z)$ must be
finite, it is not impossible that the temperature vanishes with the
mass in models of the DSR1 class. Therefore, in comparison with the
MS proposal, our proposal of a canonical implementation of DSR leads
to a richer variety of options for the asymptotic vanishing of the
modified temperature.

We can study the behavior of the modified temperature in the DSR
models considered in the Appendix. In the DSR2 model, the
temperature coincides with $T_0$ for the MS proposal, whereas
$T_{BH}=E_P^2/[4\pi(\pi\lambda E_P^2+2m)]$ for our proposal. In this
latter case, although the temperature does not vanish for zero mass,
the situation is much better than in standard general relativity,
because $T_{BH}$ tends to the constant $1/(4\pi^2\lambda)$. On the
other hand, using the expression of the function $H$ obtained in the
Appendix for the DSR3 model (Einstein-Rosen gravitational waves) we
get $T_{BH}=[1-e^{-\pi \lambda E_P^2/(2m)}]/(4\pi^2\lambda)$ for the
MS proposal. Thus, the temperature tends also to a constant in the
limit of vanishing black hole mass. For our alternative proposal,
$T_{BH}= E_P^2 e^{-\pi \lambda E_P^2/(2m)}/(8\pi m)$ and the
temperature does indeed vanish when $m\rightarrow0$. Finally, in the
case of the DSR1 model, the temperature suffers no modifications
when the conditions that allow the application of our analysis are
satisfied [see end of Subsec. IV B and Eq. (\ref{hdsr1})].

\subsection{Derivation from the surface gravity}

Expression (\ref{temperature}) for the modified temperature and our
subsequent discussion are only applicable if the function $H$
satisfies certain conditions spelled out in Subsec. IV B. These
conditions allow us to pass from a set of inequalities involving
$H(\epsilon,\Delta\Pi)$ for a whole range of values of the auxiliary
energy-momentum to the single inequality (\ref{areacota}). This
latter inequality leads to a generalized second law that depends
only on the area of the black hole and the entropy of the ordinary
matter. For arbitrary DSR theories, however, the conditions on $H$
will not be fulfilled. Although one can find alternative conditions
on $H$ that allow to arrive at Eq. (\ref{areacota}), it is possible
to generalize the study of the modification of the temperature and
the entropy in the following way.

First, in the spirit of the gravity's rainbow \cite{rainbow}, one
can tentatively admit a black hole temperature ${\cal T}_{BH}$ that
depends on the energy-momentum of the test particles. Then, as
pointed out by Ling, Li, and Zhang \cite{ling2}, the expression for
the corresponding temperature can be derived by analyzing the
behavior of the gravity's rainbow metric near the horizon. The
temperature is ${\cal T}_{BH}=\kappa/(2\pi)$, where $\kappa$ is the
surface gravity on the horizon:
\begin{equation}\kappa=-\frac{1}{2}\lim_{r\rightarrow
r_s}\sqrt{\frac{-g^{rr}}{g^{00}}}\frac{(g^{00})^{\prime}}{g^{00}}.
\end{equation} The prime denotes the derivative with respect to $r$.
For the Schwarzschild solution, the introduction of the gravity's
rainbow results in the time and spatial scalings (\ref{GRS}), which
produce the following transformation of the metric components:
\begin{equation}
g^{00}\rightarrow \frac{g^{00}}{[G(\epsilon)]^2}, \quad\quad
g^{rr}\rightarrow\frac{g^{rr}}{[F(\epsilon,\Pi)]^2}.
\end{equation}

With these transformation laws, one straightforwardly obtains
\begin{equation}{\cal T}_{BH}=H(\epsilon,\Pi)T_0.\end{equation} This
expression depends on the energy and momentum of the test particles.
It seems reasonable to consider as natural test particles those
provided by the black hole itself by means of its radiation. In this
way, the gravity's rainbow would take into account (to a certain
extent) the back reaction of the geometry. This radiation would be
dominated by massless particles with average energy proportional to
the Hawking temperature, at least for sufficiently large black
holes. This would justify identifying the test particles as massless
ones with $\epsilon=\Pi=\xi T_0$, where $\xi$ is a constant
approximately of order unity.

Nonetheless, one should expect this to be only an approximation,
with quantum corrections to the value of the average energy and
fluctuations around the typical test particle becoming increasingly
important for smaller black holes. One could try to mimic the effect
of those corrections and departures from the proposed approximation
by evaluating $H$ at $\epsilon=\xi T_0\{1+O[(T_0/E_P)^{n_1}]\}$ and
$\Pi= \xi T_0\{1+O[(T_0/E_P)^{n_2}]\}$, with $n_1$ and $n_2$ two
positive constants and the symbol $O$ denoting the order of the
uncontrolled terms. The resulting temperature could then be
interpreted as the genuine modified temperature of the black hole,
$T_{BH}$. Assuming that $H$ is analytic in the region of small
arguments and expanding it around $\xi T_0$, one would obtain
\begin{equation}
T_{BH}= T_0+ [H(\xi T_0,\xi
T_0)-1]T_0\left\{1+O\left[\frac{T_0^{n}}{E_P^{n}}\right]\right\}.
\end{equation}
We have used that $H(0,0)=1$. Here, $n$ is the minimum of $n_1$ and
$n_2$ and may in principle be any positive constant. As a
consequence, in the above expression one would generally be sure
only of the significance of the leading order correction to the
Hawking temperature, correction that arises from the first
nonconstant term in the Taylor series of $H$ around zero. This
contrasts with the situation found for theories that satisfy the
requirements introduced in Subsec. IV B, for which a full expression
for the modified temperature has been derived.

\section{Summary and Conclusions}

We have studied the modified entropy and temperature of
(Schwarzschild) black holes in the framework of a gravity's rainbow.
We have considered two different formalisms of this type. One is the
original gravity's rainbow proposed by Magueijo and Smolin. The
other is a related formalism based on a proposal of the authors
about a canonical implementation of DSR. This implementation leads
to a set of spacetime coordinates that are canonically conjugate to
the physical energy and momentum. In both formalisms, the metric
depends (explicitly) on the energy and momentum of the particle that
is supposed to test the geometry. We have discussed the implications
that both modified dispersion relations and generalized uncertainty
principles have on black hole thermodynamics. Gravity's rainbow
formalisms incorporate these two kinds of modifications and, in this
sense, provide a suitable arena to carry out the desired discussion.
In this context, we have focused our attention on the rather general
case when the DSR theory associated with the gravity's rainbow has a
physical energy that depends only on the undistorted energy of
standard special relativity (and the physical momentum vanishes if
the undistorted one does).

As starting point for our discussion we have employed Bekenstein's
calculation about the minimum change of area that a black hole
suffers when it absorbs a particle. More specifically, we have taken
into account the quantum nature of the particle in that calculation.
For both the MS proposal and our proposal, we have motivated the
introduction of a modified lower bound on the change of area that
reproduces the one derived by Bekenstein in general relativity,
except for a factor that depends on the (undistorted)
energy-momentum of the absorbed particle. We have shown that, for a
certain set of DSR theories, the different bounds obtained with the
allowed range of energies and momenta for the particle can be
captured in a single bound whose corrective factor is just a
function of the black hole area. For such DSR theories, the
energy-momentum dependent factor is an increasing function of the
momentum but becomes a decreasing function once its two arguments
(energy and momentum) are made to coincide. In particular, these
conditions of monotony are satisfied by the DSR3 analogue of the
Einstein-Rosen waves and by the familiar representative of the DSR2
models. For the usual representative of the DSR1 models the
situation is more complicated; only with our proposal for a
canonical implementation of DSR and under certain circumstances, the
model satisfies the commented conditions.

In addition, we have explored the consequences of adopting two
different types of quantization for the system. In one of them, the
elementary position and momentum variables are those associated with
flat space, and the time of the flat background is taken as the
evolution parameter of the quantum dynamics. In the other case, the
elementary position and momentum variables are the physical
variables of the DSR theory, and the evolution is given in terms of
the physical time of the system. At least as far as the analyzed
bound on the change of black hole area (and the corresponding
modified temperature) is concerned, we have shown that the former of
these quantum descriptions is connected with our canonical
formalism, whereas the latter is related to the MS proposal.

By employing the modified bound obtained for the area change
together with the Bekenstein bound for the entropy-to-energy ratio,
we have derived an inequality that can be interpreted as a
(modified) generalized second law of thermodynamics. In this manner,
we have been able to identify a modified expression for the black
hole entropy, given as a function of the area. Using this expression
we have deduced the modified temperature of the black hole. We have
shown that, for the gravity's rainbow of Magueijo and Smolin, this
temperature can vanish in the limit of zero black hole mass only in
the case of some particular DSR3 models. With our proposal for a
canonical implementation, on the other hand, the temperature may
vanish for a much ampler family of models. In particular, models of
the three distinct types of DSR families are allowed. This result
suggests that black holes might stop their evaporation or expend an
infinite time in the process, opening an avenue for the resolution
of the information loss problem in black hole physics. This issue
deserves further research along the lines that have already been
proposed in Refs. \cite{amelinoBH2,ling1}.

Our analysis provides a systematic elaboration of the arguments
based on the Bekenstein bound for its application to DSR theories
and gravity's rainbow formalisms, assuming the validity of these
formalisms as extensions of DSR that incorporate the effect of
curvature \cite{rainbow}. An important influence must be attributed
to Refs. \cite{amelinoBH1,amelinoBH2,ling1,ling2}. Comparing our
study with them, we have extended the discussions that were
available in the literature to DSR models whose physical momentum
depends nonlinearly on the auxiliary one corresponding to special
relativity. In addition, our study covers not only the MS proposal
for a gravity's rainbow, but also a proposal whose distinctive
feature is the canonical implementation of DSR
\cite{nuestro,nuestro2}. As we have seen, the modified black hole
thermodynamics arising from this alternative proposal has very
appealing physical properties, nicer in general than those deduced
with the MS construction.

The black hole metrics that we have considered are solutions to the
modified Einstein equations that arise in the gravity's rainbow
formalisms obtained with these two different proposals. In both
cases, the geometry and the gravitational constant generally depend
on the energy-momentum of the test particle that is used as a probe
by the observer: employing the language of renormalization theory,
geometry ``runs'' \cite{rainbow}. Nonetheless, it is worth
clarifying that, for each given energy-momentum, these modified
equations possess the same invariance under changes of coordinates
as in general relativity. In particular, instead of using
coordinates of the Schwarzschild type like in Ref. \cite{rainbow},
one can describe the modified Schwarzschild solutions in any other
set of coordinates (e.g. the generalization of the
Eddington-Finkelstein or the Kruskal-Szekeres coordinates) without
changing the conclusions. Similarly, for each energy-momentum of the
test particle, one can introduce well-defined notions of spatial and
null infinity, showing that the black hole solution is asymptotic
flat \cite{hackett}.

Finally, an interesting line of research for future investigations
is the possible existence of {\it modified} black hole analogues in
condensed matter physics. It is known that, under certain
approximations, the description of some condensed matter systems can
be split into a background configuration, interpretable as a black
hole geometry, and some relativistic fields propagating on it
\cite{analog}. Beyond the geometric regime in which these
approximations are valid, the relativistic fields adopt modified
dispersion relations, with corrections that become important for
large frequencies. This situation presents a suggestive parallelism
with that encountered in the gravitational theories that we have
considered. In this respect, the fact that the underlying physics
and the emergence of modifications are well understood in analogue
models can be an important plus.

\acknowledgments

The authors want to thank L.J. Garay and G. Jannes for fruitful
conversations and enlightening discussions. P.G. is also thankful to
J.M. Mart\'{i}n-Garc\'{i}a and J. Cortez for their valuable help.
P.G. gratefully acknowledges the financial support provided by the
I3P framework of CSIC and the European Social Fund. This work was
supported by funds provided by the Spanish MEC Projects No.
FIS2004-01912 and FIS2005-05736-C03-02.

\section*{Appendix: DSR models}
\setcounter{equation}{0}
\renewcommand{\theequation}{A.\arabic{equation}}

In this Appendix, we give details about three specific DSR models
that have been analyzed in the literature.

\textbf{DSR2}. The first of these models is taken as the prototype
of the so-called DSR2 theories, in which both the physical energy
and the physical momentum present an upper bound. In this model the
nonlinear action of the Lorentz group in momentum space is generated
by combining each boost with a dilatation. The model is
characterized by the functions \cite{DSR2,DSR2b}
\begin{equation}\label{gf2}
g(\epsilon)=\frac{\epsilon}{1+\lambda_2\epsilon},\quad\quad
f(\epsilon,\Pi)=\frac{\Pi}{1+\lambda_2\epsilon}.
\end{equation}
For the two rainbow realizations analyzed in the main text, namely,
the MS proposal and our proposal for a canonical implementation of
DSR (that we will call the canonical proposal in the rest of this
Appendix), expressions (\ref{gf2}) lead to the following functions
$F,$ $G$, and $H$:

i) MS proposal:
\begin{eqnarray}
G(\epsilon)&=&\frac{g(\epsilon)}{\epsilon}=
\frac{1}{1+\lambda_2\epsilon},\nonumber\\
F(\epsilon,\Pi)&=&\frac{f(\epsilon,\Pi)}{\Pi}=
\frac{1}{1+\lambda_2\epsilon},\nonumber\\
H(\epsilon,\Pi)&=&\frac{G(\epsilon)}{F(\epsilon,\Pi)}=1.
\end{eqnarray}

ii) Canonical proposal:
\begin{eqnarray}
G(\epsilon)&=&\partial_\epsilon g(\epsilon)
=\frac{1}{(1+\lambda_2\epsilon)^2} \nonumber\\
F(\epsilon,\Pi)&=&\partial_\Pi f(\epsilon,\Pi)=
\frac{1}{1+\lambda_2\epsilon},\nonumber\\
H(\epsilon,\Pi)&=&\frac{G(\epsilon)}{F(\epsilon,\Pi)}
=\frac{1}{1+\lambda_2\epsilon}.\end{eqnarray}

\textbf{DSR3}. The second example is a DSR version of the
Einstein-Rosen waves. These are vacuum solutions to general
relativity that describe cylindrical gravitational waves with linear
polarization (i.e., spacetimes with an axial spacelike Killing
vector and a translational one that commute and are hypersurface
orthogonal). The connection between cylindrical gravity and DSR has
been analyzed in Ref. \cite{erdsr}. For these waves, the physical
energy turns out to be given by a nonlinear function of a different,
auxiliary energy that is defined via quantum field theory in flat
spacetime \cite{ER}. For each angular frequency and wavenumber
$(\epsilon,\Pi)$ in this auxiliary theory, the nonlinear relation is
\begin{equation}
g(\epsilon)=\frac{1-e^{-\lambda_3\epsilon}}{\lambda_3},\quad\quad
f(\Pi)=\Pi.
\end{equation}This model can be regarded as a DSR3 theory, with
bounded physical energy but unbounded momentum. The corresponding
functions $F,$ $G$, and $H$ have the following form:

i) MS proposal:
\begin{eqnarray}G(\epsilon)&=&\frac{g(\epsilon)}{\epsilon}
=\frac{1-e^{-\lambda_3\epsilon}}{\lambda_3\epsilon},\nonumber\\
F(\epsilon,\Pi)&=&\frac{f(\epsilon,\Pi)}{\Pi}=1,\nonumber\\
H(\epsilon,\Pi)&=&\frac{G(\epsilon)}{F(\epsilon,\Pi)}
=\frac{1-e^{-\lambda_3\epsilon}}{\lambda_3\epsilon}.\end{eqnarray}

ii) Canonical proposal:
\begin{eqnarray}G(\epsilon)&=&\partial_\epsilon g(\epsilon)=
e^{-\lambda_3\epsilon},\nonumber\\
F(\epsilon,\Pi)&=&\partial_\Pi f(\epsilon,\Pi)=1,\nonumber\\
H(\epsilon,\Pi)&=&\frac{G(\epsilon)}{F(\epsilon,\Pi)}
=e^{-\lambda_3\epsilon}.\end{eqnarray}

\textbf{DSR1}. The third example was actually the first DSR model
that appeared in the literature \cite{DSRi}. For this reason, all
models that share with it the property of possessing a bounded
physical momentum but unbounded physical energy are said to belong
to the DSR1 class. The functions that determine this model are
\cite{DSRi,DSRii,judes}
\begin{eqnarray}
g(\epsilon|\eta)&=&\frac{1}{\lambda_1}\ln\left[1+\lambda_1\epsilon
\sqrt{1+\frac{\lambda_1^2\eta^2}{4}}
+\frac{\lambda_1^2\eta^2}{2}\right],\nonumber\\
\nonumber\\
f(\epsilon,\Pi|\eta)&=&
\frac{\Pi\sqrt{1+\frac{\lambda_1^2\eta^2}{4}}}
{1+\lambda_1\epsilon \sqrt{1+\frac{\lambda_1^2\eta^2}{4}}
+\frac{\lambda_1^2\eta^2}{2}},\label{fg1}
\end{eqnarray}
where $\eta^2:=\epsilon^2-\Pi^2$ is the Casimir invariant.

We restrict our attention to the case of fixed $\eta$, because the
function $g$ would otherwise depend on the auxiliary momentum,
contradicting the assumptions of our analysis in the main text.
Moreover, in the case of the MS proposal, we will further
concentrate our study exclusively on massless particles (e.g.,
photons) in order to simplify the discussion. Substituting $\eta=0$
directly in Eq. (\ref{fg1}), one obtains
\begin{eqnarray}
g_0(\epsilon)&:=&g(\epsilon|\eta=0)=\frac{1}{\lambda_1}
\ln\left(1+\lambda_1\epsilon\right),\nonumber\\
f_0(\epsilon,\pi)&:=&f(\epsilon,\Pi|\eta=0)=
\frac{\Pi}{1+\lambda_1\epsilon}.
\end{eqnarray} Using these functions for the MS proposal and, more
generally, the functions (\ref{fg1}) for our canonical proposal, it
is a simple exercise to deduce the following expressions for $F,$
$G$, and $H$:

i) MS proposal for massless particles:
\begin{eqnarray}G(\epsilon)&=&\frac{g_0(\epsilon)}{\epsilon}
=\frac{\ln\left(1+\lambda_1\epsilon\right)}
{\lambda_1\epsilon},\nonumber\\
F(\epsilon,\Pi)&=&\frac{f_0(\epsilon,\Pi)}{\Pi}
=\frac{1}{1+\lambda_1\epsilon}, \nonumber\\
H(\epsilon,\Pi)&=&\frac{G(\epsilon)}{F(\epsilon,\Pi)}
=\frac{1+\lambda_1\epsilon}{\lambda_1\epsilon}
\ln\left(1+\lambda_1\epsilon\right).\end{eqnarray}

ii) Canonical proposal:
\begin{eqnarray}G(\epsilon)&=&\partial_\epsilon g(\epsilon|\eta)=
\frac{\sqrt{1+\frac{\lambda_1^2\eta^2}{4}}} {1+\lambda_1\epsilon
\sqrt{1+\frac{\lambda_1^2\eta^2}{4}} +\frac{\lambda_1^2\eta^2}{2}},
\nonumber\\
F(\epsilon,\Pi)&=&\partial_\Pi f(\epsilon,\Pi|\eta)
=\frac{\sqrt{1+\frac{\lambda_1^2\eta^2}{4}}} {1+\lambda_1\epsilon
\sqrt{1+\frac{\lambda_1^2\eta^2}{4}} +\frac{\lambda_1^2\eta^2}{2}},
\nonumber\\
H(\epsilon,\Pi)&=&\frac{G(\epsilon)}{F(\epsilon,\Pi)}=1.
\label{hdsr1}\end{eqnarray}

The invariant scales $\lambda_n$ (with $n=1,2,3$) for the different
DSR models do not necessarily coincide. Nonetheless, we have
obviated this difference in the main text for simplicity, adopting
the notation $\lambda$ for all of them.


\begin{thebibliography}{60}

\bibitem{bek} J.D. Bekenstein, Phys. Rev. D {\bf 7}, 2333 (1973);
Lett. Nuovo Cimento {\bf 4}, 737 (1972); Ph.D. thesis, Princeton
University, 1972 (unpublished).

\bibitem{hod} S. Hod, Phys. Rev. D {\bf 59}, 024014 (1999).

\bibitem{bekstring} A. Strominger and C. Vafa, Phys. Lett. B
{\bf 379}, 99 (1996).

\bibitem{bekloop} C. Rovelli, Phys. Rev. Lett. {\bf 77}, 3288
(1996); A. Ashtekar, J. Baez, A. Corichi, and K. Krasnov, Phys. Rev.
Lett. {\bf 80}, 904 (1998).

\bibitem{logstring} S.N. Solodukhin, Phys. Rev. D {\bf 51}, 609
(1995); {\bf 51}, 618 (1995); {\bf 57}, 2410 (1998).

\bibitem{logloop} R.K. Kaul and P. Majumdar, Phys. Rev. Lett.
{\bf 84}, 5255 (2000); A. Chatterjee and P. Majumdar, Phys. Rev.
Lett. {\bf 92}, 141301 (2004); Pramana {\bf 63}, 851 (2004).

\bibitem{loggeneral} R.J. Adler, P. Chen, and D.I. Santiago,
Gen. Rel. Grav. {\bf 33}, 2101 (2001).

\bibitem{chen} P. Chen and R.J. Adler, Nucl. Phys. Proc. Suppl. {\bf
124}, 103 (2003).

\bibitem{medved} M. Cavaglia, S. Das, and R. Maartens, Classical
Quantum Gravity {\bf 20}, L205 (2003); M. Cavaglia and S. Das,
Classical and Quantum Gravity {\bf 21}, 4511 (2004); A.J.M. Medved
and E.C. Vagenas, Phys. Rev. D {\bf 70}, 124021 (2004).

\bibitem{amelinoBH1} G. Amelino-Camelia, M. Arzano, and A.
Procaccini, Phys. Rev. D {\bf 70}, 107501 (2004); Int. J. Mod. Phys.
D {\bf 13}, 2337 (2004).

\bibitem{amelinoBH2} G. Amelino-Camelia, M. Arzano, Y. Ling, and G.
Mandacini, Classical Quantum Gravity {\bf 23}, 2585 (2006).

\bibitem{ling1} Y. Ling, B. Hu, and X. Li, Phys. Rev. D
{\bf 73}, 087702 (2006).

\bibitem{ling2} Y. Ling, X. Li, and H. Zhang, gr-qc/0512084.

\bibitem{MDRloop} R. Gambini and J. Pullin, Phys. Rev. D {\bf 59},
124021 (1999).

\bibitem{MDRloop2} J. Alfaro, H.A. Morales-Tecotl, and L.F.
Urrutia, Phys. Rev. Lett. {\bf 84}, 2318 (2000); Phys. Rev. D {\bf
65}, 103509 (2002).

\bibitem{MDRloop3} L. Smolin, Nucl. Phys. B {\bf 742}, 142
(2006); arXiv: hep-th/0209079.

\bibitem{MDRnoconm} J. Lukierski, A. Nowicki, and H. Ruegg, Phys.
Lett. B {\bf 293}, 344 (1992).

\bibitem{GUPstring} G. Veneziano, Europhys. Lett. {\bf 2}, 199
(1986); D. Amati, M. Ciafaloni, and G. Veneziano, Phys. Lett. B {\bf
197}, 81 (1987); Int. J. Mod. Phys. A {\bf 3}, 1615 (1988).

\bibitem{GUPstring2} D.J. Gross and P.F. Mende, Phys. Lett. B {\bf
197}, 129 (1987); Nucl. Phys. B {\bf 303}, 407 (1988).

\bibitem{GUPstring3} T. Yoneya, Mod. Phys. Lett. A {\bf 4}, 1587
(1989); K. Konishi, G. Paffuti, and P. Provero, Phys.
Lett. B {\bf 234}, 276 (1990); R. Guida, K. Konishi, and P. Provero,
Mod. Phys. Lett. A {\bf 6}, 1487 (1991).

\bibitem{GUPnoconm} S. Doplicher, K. Fredenhagen, and J.E. Roberts,
Phys. Lett. B {\bf 331}, 39 (1994).

\bibitem{GUP0} M. Maggiore, Phys. Lett. B {\bf 304}, 65 (1993);
Phys. Rev. D {\bf 49}, 5182 (1994); F. Scardigli, Phys. Lett. B {\bf
452}, 39 (1999); R.J. Adler and D.I. Santiago, Mod. Phys. Lett. A
{\bf 14}, 1371 (1999).

\bibitem{fenom} G. Amelino-Camelia, J.R. Ellis, N.E. Navromatos,
D.V. Nanopoulos, and S. Sarkar, Nature (London) {\bf 393}, 763
(1998).

\bibitem{fenom2} G. Amelino-Camelia and T. Piran, Phys. Lett. B {\bf
497}, 265 (2001); Phys. Rev. D {\bf 64}, 036005 (2001).

\bibitem{fenom3} G. Amelino-Camelia, Phys. Lett. B {\bf 528}, 181
(2002); Mod. Phys. Lett. A {\bf 17}, 899 (2002).

\bibitem{fenom4} S. Sarkar, Mod. Phys. Lett. A {\bf 17}, 1025
(2002).

\bibitem{fenom5} T. Jacobson, S. Liberati, and D. Mattingly, Phys.
Rev. D {\bf 66}, 081302 (2002).

\bibitem{DSRi} G. Amelino-Camelia, Int. J. Mod. Phys. D {\bf 11},
35 (2002).

\bibitem{DSRii} G. Amelino-Camelia, Phys. Lett. B {\bf 510}, 255
(2001).

\bibitem{DSR1} J. Kowalski-Glikman, Phys. Lett. A {\bf 286}, 391
(2001).

\bibitem{DSR1b} N.R. Bruno, G. Amelino-Camelia, and J.
Kowalski-Glikman, Phys. Lett. B {\bf 522}, 133 (2001).

\bibitem{DSR2} J. Magueijo and L. Smolin, Phys. Rev. Lett. {\bf 88},
190403 (2002).

\bibitem{DSR2b} J. Magueijo and L. Smolin, Phys. Rev. D {\bf 67},
044017 (2003).

\bibitem{DSR12} J. Kowalski-Glikman and S. Nowak, Phys. Lett. B
{\bf 539}, 126 (2002).

\bibitem{Luis} L.J. Garay, Int. J. Mod. Phys. A {\bf 10}, 145
(1995); and references therein.

\bibitem{rainbow} J. Magueijo and L. Smolin, Classical Quantum
Gravity {\bf 21}, 1725 (2004).

\bibitem{nuestro} P. Gal\'an and G.A. Mena Marug\'an, Phys. Rev.
D {\bf 70}, 124003 (2004).

\bibitem{nuestro2} P. Gal\'an and G.A. Mena Marug\'an, Phys. Rev.
D {\bf 72}, 044019 (2005).

\bibitem{christ} D. Christodoulou, Phys. Rev. Lett. {\bf 25},
1596 (1970); Ph.D. thesis, Princeton University, 1971 (unpublished);
D. Christodoulou and R. Ruffini, Phys. Rev. D {\bf 4}, 3552 (1971).

\bibitem{note1} Ultimately, $L$ is the characteristic length
of the falling system.

\bibitem{cotaS/E} J.D. Bekenstein, Phys. Rev. D {\bf 23}, 287
(1981).

\bibitem{hawking} S.W. Hawking, Comm. Math. Phys. {\bf 43}, 199
(1975).

\bibitem{judes} S. Judes and M. Visser, Phys. Rev. D {\bf 68}, 045001
(2003).

\bibitem{DSRposition} J. Lukierski, hep-th/0402117.

\bibitem{DSRposition2} D. Kimberly, J. Magueijo, and J. Medeiros,
Phys. Rev. D {\bf 70}, 084007 (2004).

\bibitem{DSRposition3} S. Mignemi, Phys. Rev. D {\bf 68},
065029 (2003).

\bibitem{hinterleitner} F. Hinterleitner, Phys. Rev. D
{\bf 71}, 025016 (2005).

\bibitem{note1b} Actually, the careful reader can check that
this restriction is not necessary for the discussion of the proposal
for a canonical implementation of DSR.

\bibitem{note2} Strictly speaking, one has
$\Delta q^i \Delta\Pi_i \geq1/2$ (for each value of $i$). The
passage from vector components to norms is valid under reasonable
assumptions.

\bibitem{note3} Conversely, the relation $A=4 \pi r_s^2$ can be
seen as the definition of $A$ in terms of the radial position of the
horizon.

\bibitem{note4} This statement implicitly assumes that the
bound $\overline{\Delta A}_0$ for standard general relativity can be
dealt with independently of $(\epsilon,\Delta\Pi)$, as indicated by
the fact that the minimum $a$ of $\overline{\Delta A}_0$ is
independent of those quantities.

\bibitem{ER} J.F. Barbero G., G.A. Mena Marug\'{a}n, and E.J.S.
Villa\-se\~{n}or, Phys. Rev. D {\bf 67}, 124006 (2003); {\bf 69},
044017 (2004).

\bibitem{erdsr} L. Freidel, J. Kowalski-Glikman, and L. Smolin,
Phys. Rev. D {\bf 69}, 044001 (2004).

\bibitem{hackett} J. Hackett, Classical Quantum Gravity {\bf
23}, 3833 (2006).

\bibitem{analog} L.J. Garay, J.R. Anglin, J.I. Cirac, and P. Zoller,
Phys. Rev. Lett. {\bf 85}, 4643 (2000); U. Leonhardt and P.
Piwnicki, Phys. Rev. Lett. {\bf 84}, 822 (2000); G.E. Volovik, Phys.
Rep. {\bf 351}, 195 (2001); {\it The Universe in a Helium Droplet}
(Oxford University Press, 2003); {\it Artificial Black Holes},
edited by M. Novello, M. Visser, and G.E. Volovik (World Scientific,
2002); C. Barcel\'{o}, S. Liberati, and M. Visser, Living Rev. Rel.
{\bf 8}, 12 (2005).

\end{thebibliography}
\end{document}